\documentclass[twocolumn,showpacs,citeautoscript,floatfix]{revtex4}
\usepackage{amssymb}
\usepackage{graphicx}
\usepackage{dcolumn}   % column centering
\usepackage{bm}        % bold math
\usepackage[]{amsmath}
\usepackage{caption}%http://ctan.org/pkg/caption
\usepackage[usenames]{color}
\usepackage[normalem]{ulem}

\begin{document}

\title{ Doping-induced spin-manipulation in complex trimer system Ca$_{3}$Cu$_{3}$(PO$_{4}$)$_{4}$: A First Principles Study}

\author{Debjani Karmakar}

\affiliation{ Technical Physics Division, Bhabha Atomic research Center, Mumbai 400085, India.}

\author{Alexander Yaresko}

\affiliation{ Max-Planck-Institut f\"{u}r Festk\"{o}rperforschung, Heisenbergstra{\ss}e 1, D-70569, Stuttgart, Germany.}

\date{\today}
\begin{abstract}
Spin-manipulated doping with magnetic (Ni) and non-magnetic (Mg) dopants constitutes the experimental attempts to obtain a singlet ground state system from the linear chain Heisenberg antiferromagnetic Cu-based $d^{9}$ spin-1/2 trimer compound Ca$_{3}$Cu$_{3}$(PO$_{4}$)$_{4}$ with doublet ground state. The present study is a first-principles based investigation of the effects of such doping on the spin-exchange mechanism and electronic structure of the parent compound. Site-selective doping with zero-spin dopants like Mg is proved to be more efficient than an integral spin dopant Ni in obtaining a spin-gap system with singlet ground state, as also observed in the experimental studies. Doping induced dimerized state is found to be the lowest in ground-state energy. Calculated spin exchange values along various possible paths resemble nicely with earlier experimental results.
\end{abstract}
\pacs{75.50.Ee, 71.15.Mb, 75.10.Jm}
\maketitle
%\yr{This text I propose to remove}\ya{and insert this one}
%\yc{And these are my comments}

\section{Introduction}
Basic concepts of spin-exchange mechanism and excited state properties of low dimensional magnetic systems can be useful to provide basic models for some complex phenomena like high-temperature superconductivity in cuprates and arsenides.\cite{vasilev} Low-dimensionality of magnetic exchange properties when associated with an energy-gap in the spin-excitation spectrum is even more interesting due to it's intrinsic two-level behaviour,\cite{ueda, bose} as demanded for utilization of such systems as a future quantum computing device material.\cite{mayaffre} The simplest of such system, single-dimensional spin-1/2 Heisenberg antiferromagnetic(HAFM)dimer chain is well-known to be deprived of long-range magnetic order.\cite{haldane1, haldane2} It's zero-spin singlet ground state is separated from the integral-spin triplet excited state with an energy gap, known as spin-gap. Above a certain critical field($h_{c}$), the Zeeman splitted triplet excited states starts populating leading to a field-induced Bose-Einstein condensation and rendering the ground-state above $h_{c}$ to be the Bose condensate of low-energy magnons.\cite{affleck}

Inelastic Neutron scattering experiments and successive analysis of exchange integrals have enabled to obtain an understanding of many such systems where the magnetic interactions within a geometric cluster of atoms are orders of magnitude larger than the inter-cluster interactions.\cite{matsuda, ruegg} The ground-state properties of such systems are very much sensitive of intra and inter cluster interactions. One such quasi-two-dimensional weakly coupled system is the triangle spin-1/2 trimer system La$_{4}$Cu$_{3}$Mo$_{3}$O$_{12}$, which can be well modelled by isolated Heisenberg spin-triangle model.\cite{azuma} Another geometrically frustrated isolated half-integral spin-trimer system is Vanadium Hallide, where the consecutive frustrated spins in an isolated triangular plaquette are aligned at 120 degrees with respect to each other.\cite{hirakawa} The zero-temperature ground state magnetic structures of all such kind of systems are very much sensitive of the intra and inter-trimer interactions.

\begin{figure*}
{\includegraphics[width=80mm,clip]{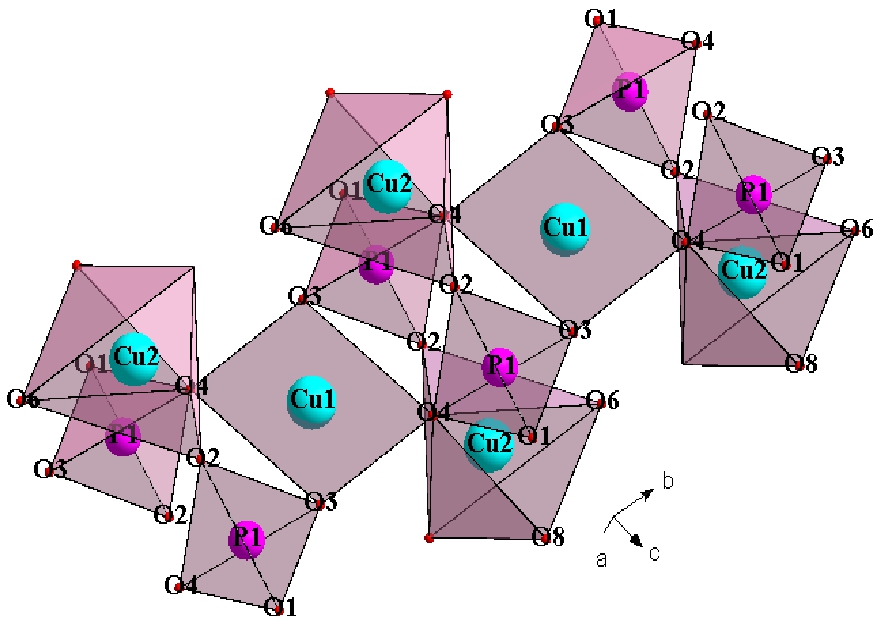}}
{\includegraphics[width=80mm,clip]{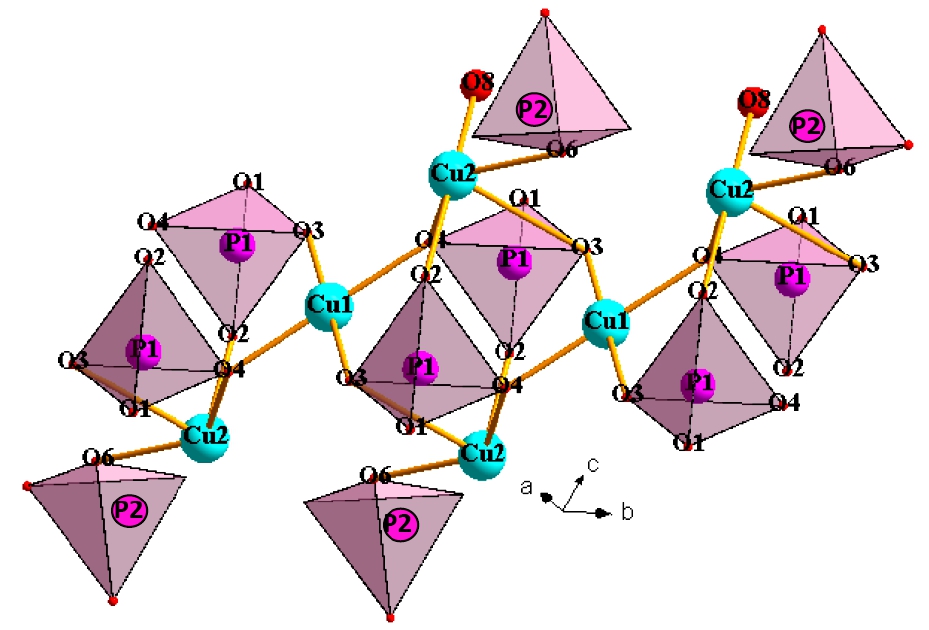}}
{\includegraphics[width=80mm,clip]{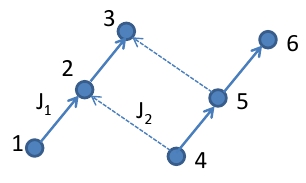}}
\caption{\label{fig1a} (colour online)(a): In a double-trimer array, the square planar and distorted square pyramidal coordination with O-ligands for Cu1 and Cu2 ions. Four corners of the Cu1-O3-O4 Square are shared with P1O$_{4}$ tetrahedra. (b): The double trimer Cu2-Cu1-Cu2 array (discussed in text). Inter-trimer exchanges are mediated via O4 of P1O$_{4}$ tetrahetra. (c) Schematic view of the significant exchange couplings in the \textit{b}-axis trimer array. $J_{intra}=J_{1}$ and $J_{inter}=J_{2}$ (see text for details) are denoted by solid and dotted lines. }
\end{figure*}

In the present study, the parent system, Ca$_{3}$Cu$_{3}$(PO$_{4}$)$_{4}$ is a model spin-1/2 HAFM spin-trimer system with two symmetric site of Cu, where Cu2-Cu1-Cu2 trimers form a linear chain along \textit{b}-axis with intra and inter-trimer exchange to be 126 \textit{K} and 3 \textit{K} respectively. \cite{matsuda, belik, pomjakushin1} The ground-state for this system will be a doublet with magnetic configuration Cu2$^{2+}(\uparrow)$-Cu1$^{2+}(\downarrow)$-Cu2$^{2+}(\uparrow)$. In principle, if Cu1$^{2+}$ can be replaced by a spin-one ion like Ni$^{2+}$, the ground state of the system may be converted into a singlet one. The experimental observation by Pomjakushin \textit{et}. \textit{al}. \cite{pomjakushin}with powder inelastic neutron scattering analysis reveals that such selective doping does not occur in reality. Stable structure with Ni substituting Cu1 belong to a different space group $C2/c$ with a doubled unit cell along \textit{c}-axis, which differs from the original space group $P2_{1}/a$ for the parent compound Ca$_{3}$Cu$_{3}$(PO$_{4}$)$_{4}$. Instead of Cu1, Ni prefers to replace the end position Cu2 of the trimer chain. Hence, Ni does not satisfy the criteria of becoming a suitable dopant to induce a singlet ground-state in the substituted compound. In a successive study, Ghosh \textit{et}. \textit{al}. could successfully obtain a singlet ground state for this compound by selectively substituting Cu2 with a non-magnetic ion Mg only at one end position of the trimer chain. \cite{ghosh}

\begin{figure*}
{\includegraphics[width=160mm,clip]{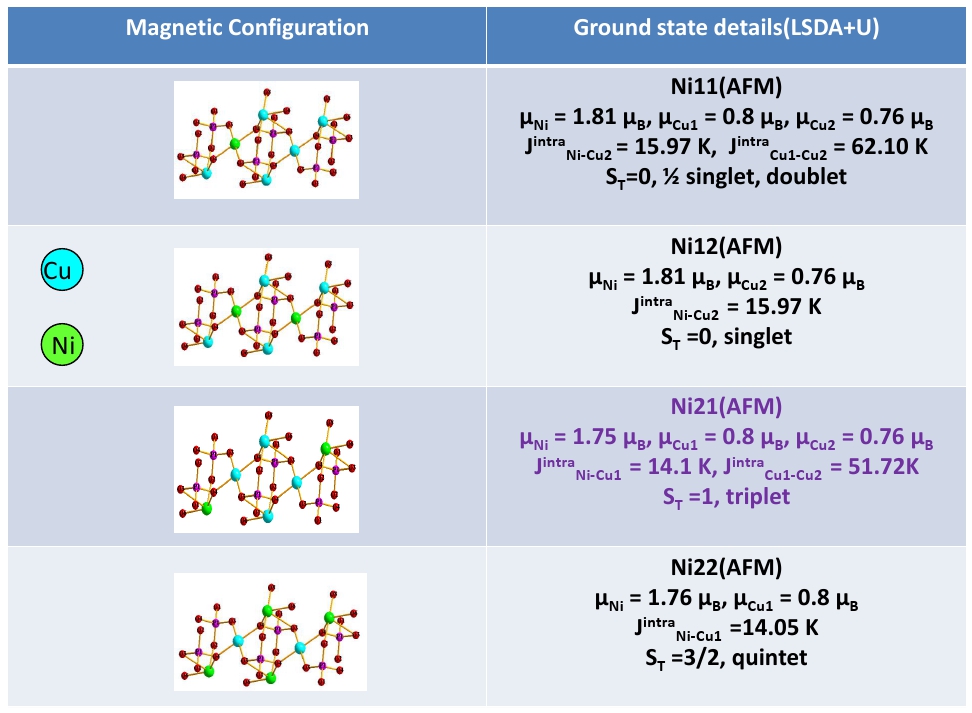}}

Table I: (colour online)Various substitutional possibilities of Ni-substitution and their corresponding ground state details resulting from a LSDA + U calculations.
\end{figure*}

These two studies have provided the motivation for the present investigation on Ca$_{3}$Cu$_{3}$(PO$_{4}$)$_{4}$ system under spin-manipulated doping with Ni and Mg, where with a detailed electronic structure calculation, we aim to understand the effects of these dopants on the spin-exchange behaviour and ground-state. Experimentally, when a system is substitutionally doped with a particular dopant, there can be various possibilities of magnetic configuration. For this particular system, only a few of such configurations will result into a singlet ground state spin-gap system. With the help of first-principles study, we have explained the modification of spin-exchange mechanism and exchanges along various possible pathways due to doping, which resembles nicely with the earlier experimental studies.\cite{ghosh, pomjakushin}

The paper is organized as follows. In section II, we explain the details of the first principles calculation, where in various subsections, the structural details and electronic structure investigations of Ni and Mg-doped systems are presented. Exchange integrals along different superexchange paths are calculated and compared with experimental values. Partial density of state (PDOS) plots and the relevant band-structural details are also studied to understand the magnetic response of the doped systems. The last section summarizes the results with a conclusion.

\section{First principles Calculation}

Ca$_{3}$Cu$_{3}$(PO$_{4}$)$_{4}$, a suitable candidate to study spin-trimer antiferromagnetism crystallizes in the monoclinic structure with space group $P2_{1}/a$. The system consists of two symmetric site for Cu, \textit{viz}. Cu1 and Cu2 having square planar and distorted square pyramidal coordination with O-ligands, as depicted in figure 1(a). The intra-trimer superexchange path between Cu1 and Cu2 (3.45 {\AA}) is mediated via O4 consisting of the phosphate tetrahedra centering P1. Inter-trimer superexchange paths Cu1-O3-Cu2, Cu1-O3-P1-O4-Cu2 and Cu1-O4-P1-O3-Cu1 are longer than Cu1-Cu2 inter-trimer direct distance(3.57 {\AA}), as can be seen from figure 1(b).

We have investigated the electronic structure of this system with the help of linear muffin-tin orbitals (LMTO)\cite{lmto1, lmto2} method in the atomic sphere approximation (ASA), with combined correction terms, within the framework of LSDA + Hubbard U approach. The Hubbard \textit{U} is added explicitly to the LSDA Hamiltonian using the methodology prescribed by Czyzyk and Sawatzky.\cite{sawatzky} For the exchange correlation potential, we have used Perdew-Wang parametrization.\cite{perdew} Self-consistency is achieved by performing calculations on a $8\times16\times12$ \textit{k}-mesh for the Brillouin zone integration. The on-site \textit{d-d} Coulomb interaction and exchange interaction parameters (\textit{U}, \textit{J}) for the transition metal (TM) ions Ni and Cu are taken to be (4.5, 0.8) eV and (5, 1) eV respectively for all of our bulk calculations. However, these parameters are optimized after comparing the theoretically obtained exchange coupling constants with the experimental results. In the successive sections, this will be discussed in detail.
\begin{figure}
{\includegraphics[width=90mm,clip]{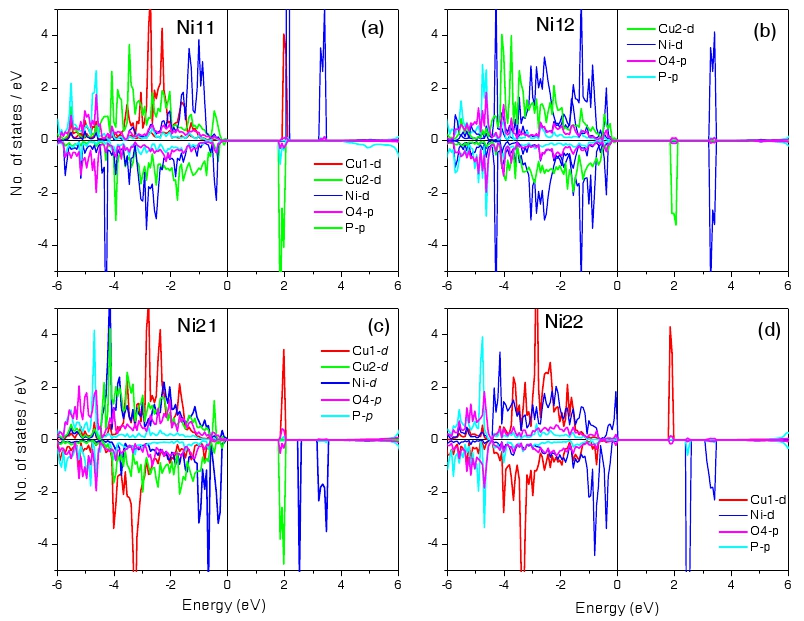}}
\caption{\label{fig3} (colour online) The PDOS for Cu1-3\textit{d}, Cu2-3\textit{d}, Ni-3\textit{d}, O4-\textit{p}, P-\textit{p} states are plotted for (a) Ni11, (b) Ni12, (c) Ni21 and (d) Ni22 resulting from a LSDA + U calculations.  }
\end{figure}

\begin{figure*}
{\includegraphics[width=160mm,clip]{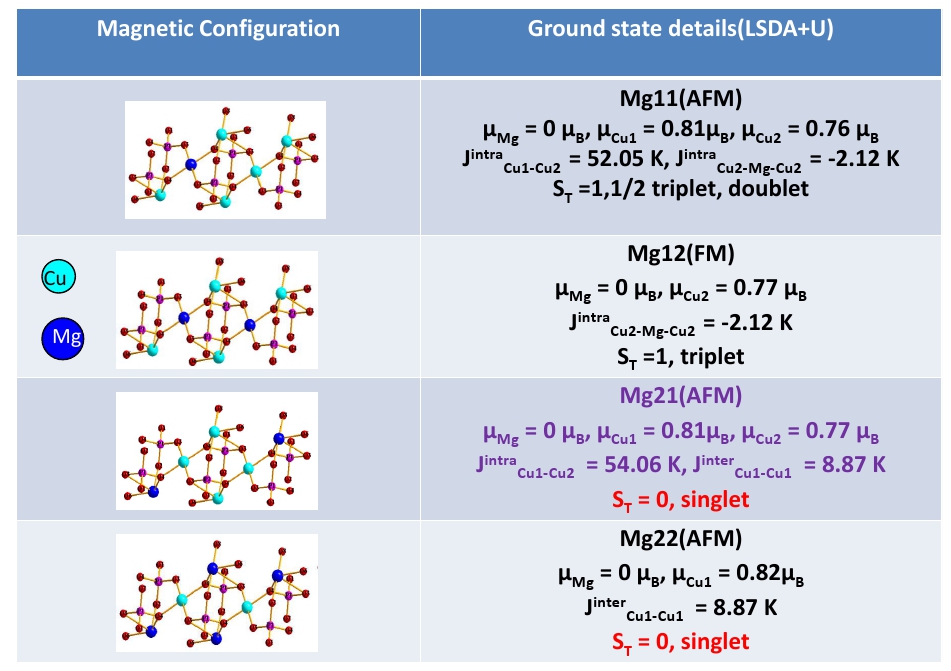}}

Table II:(colour online) Various substitutional possibilities of Mg-substitution and their corresponding ground state details resulting from a LSDA + U calculations.
\end{figure*}

To obtain a double-trimer array along \textit{b}-axis, as seen in figure 1(b), we have constructed a $1\times3\times1$ supercell to calculate the intra and inter-trimer exchanges. Realistically, in such an array,  experimental substitution with Ni and Mg may lead to partial or complete substitution at various Cu1 and Cu2 sites. In the present study, we investigate all possible magnetic configurations resulting from substitution of Ni or Mg in the parent system. We denote these feasible configurations by Nijk, where \textit{j} is the site position of Ni-doping and \textit{k} is the number of such sites replaced with Ni per trimer. Thus, in case of Ni-substitution, there can be four substitutional magnetic configurations of the double-trimer array in Figure 1(b), \textit{viz}., (a) Ni11 - substitution at a single Cu1 of one of the trimer, (b) Ni12 - substitution at Cu1 of both the trimers, (c) Ni21 - substitution at a single Cu2 of the trimer, (d) Ni22 - substitution at Cu2 of both the trimers. For Mg- substitution, the corresponding configurations are named as Mg11, Mg12, Mg21 and Mg22 respectively. In the next subsections, we will describe the ground state possibilities, corresponding intra and inter-trimer exchanges and the related electronic structure implications of all such configurations.

The model Hamiltonian for this particular system consists of two types of exchange couplings, viz. $J_{intra}=J_{1}$ and $J_{inter}=J_{2}$. Simplified Heisenberg Hamiltonian for this particular system can be written as:
\begin{eqnarray}
H=J_{1}\sum_{i=1}^{n}\textbf{S}_{3i-1}.(\textbf{S}_{3i-2}+\textbf{S}_{3i})+\notag\\
J_{2}\sum_{i=1}^{n}(\textbf{S}_{3i-1}.\textbf{S}_{3i+1}+\textbf{S}_{3i}.\textbf{S}_{3i+2})
\end{eqnarray}
Here \textit{n} is the number of trimers in an array. Schematic diagram of the trimer array is presented in figure 1(c), where different TM-spins are numbered according to the model Hamiltonian.

\subsection{Substitutional doping with Ni: resulting systems}

For the undoped parent compound, inter-trimer Cu1-Cu2 exchange, as resulted from a LSDA + U calculation is $\sim$ 92 K with the inter-trimer value to be $\sim$ 5 K. For the Ni-substituted compound, among the four possibilities, experimental investigations by Pomjakushin \textit{et}. \textit{al}. \cite{pomjakushin} have discarded Ni12, since the composition with all Cu1 sites substituted with Ni crystallizes in a different space-group $C2/c$ with a doubled unit cell along \textit{c}-axis. The symmetry analysis presented at reference \cite{pomjakushin} implies that both Cu and Ni spins can be treated in a collinear arrangement. In addition, experimental observation indicates that Ni is seen to preferentially substitute Cu2 at the end positions of the trimer. We have employed LSDA + U method for computing the total energy and exchange coupling constants, since the magnetic properties of such system are totally driven by highly localized TM-3\textit{d} states at Fermi-level. Computation of total energy by  LSDA + U calculation results into stabilization of AFM ground state with all adjacent spins in antiparallel combination for all the four Ni-substituted configurations. Interestingly, Ni12 and Ni21 is found to be the highest and lowest among them in ground-state energy. It may be mentioned in passing that by LSDA calculation, the lowest energy configuration is Ni11 with a ferromagnetic (FM) spin-alignment. Table I pictorially represents various above-mentioned configurations and their corresponding ground-state details resulting from LSDA + U calculation. Exchange coupling constant along a specified path is computed by equating the ground-state energy difference of the anti-parallel and parallel spin combination along that path with the energy differences calculated from the model Hamiltonian (equation (1)) with relevant terms incoporating individual calculated spin-values for the magnetic ions as mentioned at the table. Total spin $S_{T}$ of each such trimer configuration is specified in the same table.

For Ni11, $S_{T}$ will be either 0 or 1/2 corresponding to a singlet or doublet ground state with zero-spin Ni-substituted trimer and spin-1/2 original one. The inter-trimer exchanges are found to be very small in comparison with the intra-trimer ones and are therefore prone to small fluctuations. Ni12, although experimentally unfeasible to obtain, may have a possibility of $S_{T}$ = 0 singlet ground state. In fact, an integral spin Ni dopant was chosen by the experimentalists keeping this configuration in mind. For Ni21 and Ni22, $S_{T}$ = 1 and 3/2 corresponding to triplet and quintet ground states respectively. The intra-trimer Ni-Cu exchange is much smaller than the Cu-Cu (intra) exchange.
\begin{figure}
{\includegraphics[width=90mm,clip]{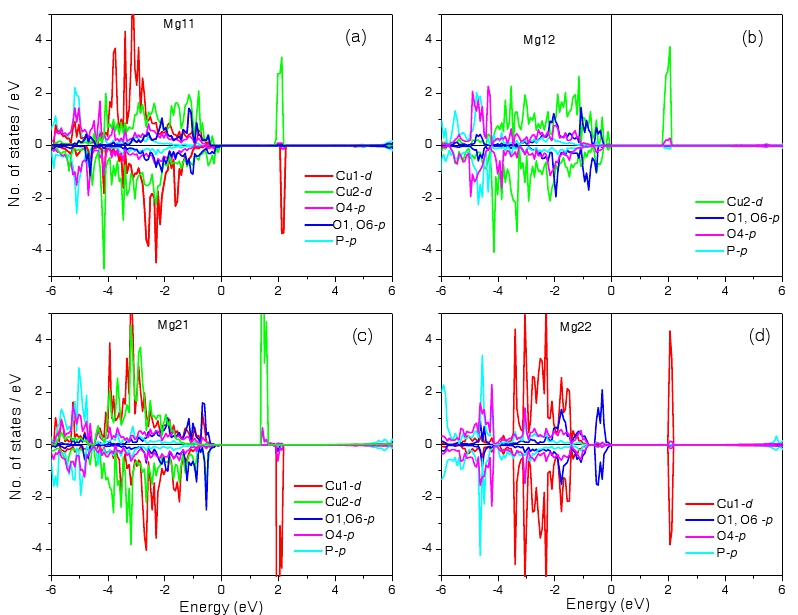}}
\caption{\label{fig5} (colour online) The PDOS for Cu1-3\textit{d}, Cu2-3\textit{d}, O4-\textit{p}, P-\textit{p} states are plotted for (a) Mg11, (b) Mg12, (c) Mg21 and (d) Mg22 resulting from a LSDA + U calculations.  }
\end{figure}
\begin{figure}
{\includegraphics[width=90mm,clip]{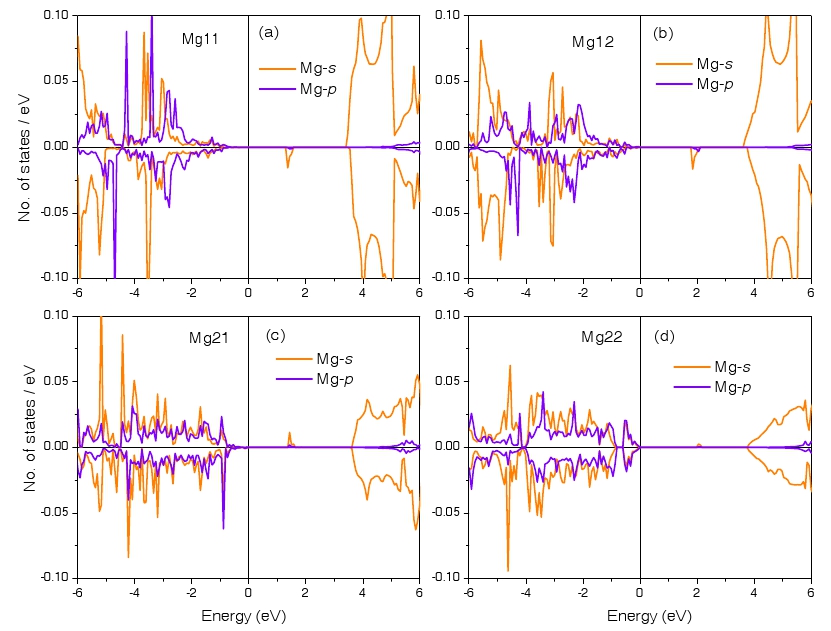}}
\caption{\label{fig6} (colour online) The PDOS for Mg-3\textit{s} and 2\textit{p} states are plotted for (a) Mg11, (b) Mg12, (c) Mg21 and (d) Mg22 resulting from a LSDA + U calculations.  }
\end{figure}

Experimental values for the Ni-Cu and Cu-Cu intra-trimer exchange was found to be 54.98 \textit{K} and 9.86 \textit{K}. Corresponding theoretical values, as computed for Ni21 (lowest energy configuration), are 51.72 \textit{K} and 14.1 \textit{K} respectively. In addition, comparing exchanges for Ni11 and Ni21 implies that, Ni, while replacing Cu2 lowers the total energy of the system more than a Cu1-replacement. Corresponding exchange couplings for Ni-Cu1 and Ni-Cu2 are 14.1 \textit{K} and 15.97 \textit{K} respectively. As expected, inter-trimer Ni-Cu and Cu-Cu exchanges are much smaller in magnitude having values of $\sim$ 1\textit{K} and 4\textit{K} respectively. To investigate the nature of magnetic exchange in more detail, we have plotted the partial density of states (PDOS) for some selective ions for all four configurations in Figure 2(a) - (d). Since, effect of O4 is the most prominent among the O-ligands, O-\textit{p} PDOS is plotted only for O4.

Available states to provide superexchange related hopping paths are determined by the square planar and square pyramidal coordination with O-ligands around Cu1 and Cu2. For all these four configurations, Cu1- and Cu2-3\textit{d} states at the up and down-spin channels have ${x^{2} - y^{2}}$ orbital-character, the orbitals being confined in the plane of O-square. For Ni11, up-spin empty states are mostly of Ni-3\textit{d}$_{x^{2} - y^{2}}$ and 3\textit{d}$_{xy}$-orbital characters, whereas for Ni21, those are of 3$d_{3z^{2} -1}$ and 3\textit{d}$_{x^{2} - y^{2}}$ orbital character. In general, Ni-3\textit{d} states are situated at a higher energy in comparison to the Cu-3\textit{d} states, leading to lowering of the value of intra-trimer Ni-Cu exchanges than Cu-Cu ones, as resulted from the theoretical calculation. Although Ni-Cu and Cu-Cu distance are of the same order, lesser probability of lowering the total spin of the system for Ni-Cu hopping leads to a decrease of the corresponding exchange couplings. Comparing figure 2(a) with 2(c), the small disparity of Ni-Cu1 and Ni-Cu2 exchanges will also be evident due to the location of the Cu1-3\textit{d}$_{x^{2} - y^{2}}$ states at slightly higher energy. Magnetic moments of Ni, Cu1 and Cu2, as calculated for Ni21 are 1.75 $\mu_{B}$, 0.8 $\mu_{B}$ and 0.76 $\mu_{B}$ respectively, whereas the corresponding experimental values are 1.8 $\mu_{B}$, 0.57 $\mu_{B}$ and 0.56 $\mu_{B}$. As mentioned in reference \cite{pomjakushin}, reduction of Cu-spins in the experimental case may be due to the frustration in the Cu-moment and due to a combination of multiple possible configurations. Magnetic moments for other configurations are mentioned in table I.
Hence, as also obtained from experimental studies, Ni-substitution in place of Cu1 or Cu2 does not lead to a singlet AFM ground state. However, the exchange coupling constants are very much dependent on the values of \textit{U} and \textit{J} parameters and therefore an optimization of the \textit{U} and \textit{J} parameters are necessary. Optimized value of U and J parameters are chosen from a comparison with the experimental value of exchanges.

\subsection{Substitutional doping with Mg: resulting systems}

Motivated by the study of reference \cite{pomjakushin}, Ghosh et. al. have succeeded in obtaining a singlet ground state by doping one non-magnetic dopant like Mg in place of Cu. Similar to Ni-substitution, experimental observation confirms the tendency of Mg to substitute the end-position (Cu2)of the trimer. In a similar way as in the previous subsection, we theoretically investigate all four different Mg-substitutional configurations Mg11, Mg12, Mg21 and Mg22 as presented in the pictorial representations of table II with an LSDA + U calculation. LSDA calculation stabilizes Mg11 as the lowest energy configuration with ferromagnetic (FM) spin-alignment.
\begin{figure}
{\includegraphics[width=60mm,clip]{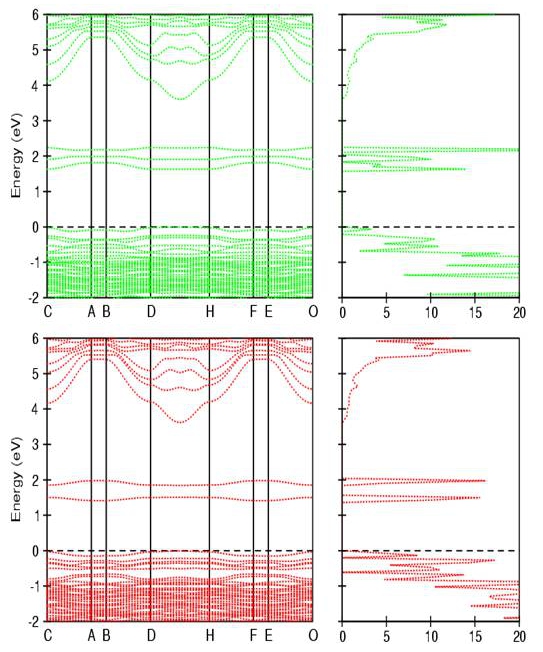}}
\caption{\label{fig6} (colour online) Band structure and the corresponding total DOS for Mg21 are plotted from a LSDA + U calculations.  }
\end{figure}
\begin{figure}
{\includegraphics[width=60mm,clip]{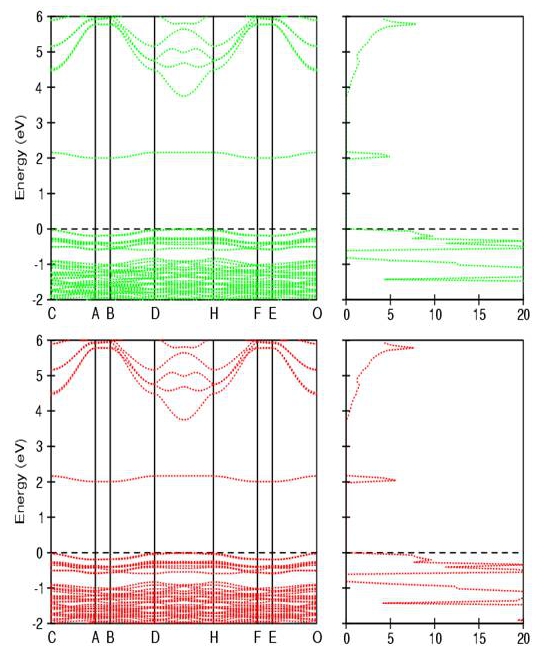}}
\caption{\label{fig6} (colour online) Band structure and the corresponding total DOS for Mg22 are plotted from a LSDA + U calculations.  }
\end{figure}

All these systems stabilize in an AFM ground-state except Mg12, where a FM ground state is dominating. Among these configurations, only Mg21 and Mg22 may lead to a singlet ground state with the former one being the lowest in ground-state energy. For Mg21, Mg substitution  transforms the trimer chain into a dimer chain. Resulting Cu1-Cu2 dimer \textit{b}-axis array leads to a reduction of the total spin of this spin-1/2 Cu2-Cu1-Cu2 trimer system and turns it into a $S_{T}$ = 0 system. Strong intra-trimer (Cu1-Cu2) and weak inter-trimer (Cu1-Cu1)interactions determine the ground state properties of Mg21. In case of  Mg22, the Cu1-spin-1/2 \textit{b}-axis array have AFM-coupled consecutive Cu1 atoms, resulting into a spin-singlet ground state. In this case, inter-trimer interaction is the only possible one. Thus with Mg-substitution, the spin-manipulation effects can be achieved in a better manner in comparison to Ni. Calculated Cu1-Cu2 intra-trimer and Cu1-Cu1 inter-trimer exchange coupling constants are both of AFM nature and estimated to be 54.06 K and 8.87 K respectively. Experimental value for the intra-trimer exchange are 55 K \cite{ghosh}.

While investigating the LSDA + U PDOS figure, presented in Figure 3(a)-(d), both for Mg11 and Mg12 in AFM and FM states, empty spin-up channel is of Cu2-3\textit{d}$_{x^{2} - y^{2}}$ character. For Mg11, empty states at spin-down channel is mostly of Cu1-3\textit{d}$_{x^{2} - y^{2}}$ character. For the lowest energy configuration Mg21 with AFM spin-arrangement, hopping to the up-spin Cu2 and down-spin Cu1-3\textit{d}$_{x^{2} - y^{2}}$ states will be more prominent. For Mg22, only Cu1-3\textit{d}$_{x^{2} - y^{2}}$ states are available for hopping. While investigating the effects of Mg-3\textit{s} and 2\textit{p}-states, we have plotted the corresponding PDOS in Figure 4(a)-(d) to understand the effects of Mg-\textit{s} and \textit{p} states on the superexchange. In case of Mg11 and Mg12, effects of Mg-\textit{s} states on the filled localized states of Cu are more, which makes the Mg bands more localized. Although very small, the Mg moments for Mg11 and Mg12 are slightly larger than those for Mg21 and Mg22. Thus, Mg-\textit{s} states have some role in mediating the superexchange for Mg11 and Mg12. In the last two cases, Mg-\textit{s} bands are less localized.  There is a substantial amount of hybridization of Mg-\textit{s} states with the P-\textit{p} states for Mg11, Mg12 and Mg21. For Mg21 and Mg22, most of the magnetic hopping is controlled by Cu1-O4-Cu2 path. Mg -\textit{s} and \textit{p} states have, therefore, a strong hybridization with O-\textit{p} and P-\textit{p} situated around Mg-O4-Cu1 path.

In Mg22, only possible superexchange interactions along inter-trimer Cu1-Cu1 array localizes the O4-2\textit{p} states. O1 and O6-2\textit{p} states are highly hybridized with Mg-3\textit{s} and 2\textit{p} states. For Mg21, O1 and O6-2\textit{p} states are less hybridized with Mg-3\textit{s} and 2\textit{p} states in comparison to  Mg22, as can be seen from figure 3(d) and 4(d).

In figures 5 and 6 , we present the LSDA + U band-structure and the corresponding total DOS for up and down-spin channels of Mg21 and Mg22, for which singlet ground states are possible. Position of highly localized empty Cu2-3\textit{d}$_{x^{2} - y^{2}}$  bands in the up-spin and Cu1-3\textit{d}$_{x^{2} - y^{2}}$  bands in the down-spin channels (as seen from Figure 5(a))for Mg21 are situated at lowest energy among the four configurations, causing the easiest lowering of energy by superexchange interactions among these four configurations. For Mg22 in an AFM configuration, both spin channels are having empty Cu1-3\textit{d}$_{x^{2} - y^{2}}$  bands. The band-gaps for Mg21 and Mg22 are $\sim$ 1.3 and 2 eV respectively. Magnetic moments of Cu1 and Cu2 ions for Mg21 are 0.81 $\mu_{B}$, 0.77 $\mu_{B}$ respectively. Theoretical analysis of energetics along with the DOS and band-structure enables to achieve a detailed understanding of the stabilization of a singlet ground state by Mg-substitution.

It may be also mentioned that structural distortion has very little to do with the magnetic exchange properties of such trimer system. In the initial case, Ni$^{2+}$ being almost of the same size as Cu$^{2+}$, does not affect the Cu2-O4-Cu1 angle much. For Ni21, Ni-O4-Cu1 angle is almost the same as Cu2-O4-Cu1 angle. For Mg21, Mg$^{2+}$ having a larger size than Cu$^{2+}$, Mg-O4-Cu1 angle is slightly larger than Cu2-O4-Cu1 angle. However this may only affect the exchange between the end points of a trimer (as in case of Mg12) and has very little effects on the dimer-chain exchange properties of the lowest energy configuration Mg21.

\section{conclusion}
In this study, we have explored the doping induced spin-manipulation of the Cu-based $d^{9}$ spin-1/2 trimer system Ca$_{3}$Cu$_{3}$(PO$_{4}$)$_{4}$ with the help of first principles LSDA + U calculation to obtain a detailed description of all possible substitutional configurations and the resulting ground states. The exchange coupling constants along various possible paths are also calculated and found to resemble nicely with experiment. The analysis of DOS and band-structure enables to understand the electronic mechanism behind obtaining a singlet ground state by substitution of Mg instead of Ni.

\section{Acknowledgement}
One of the authors (DK) would like to acknowledge the Indo-EU project MONAMI and BARC-ANUPAM supercomputing facility.

\end{document}